\documentstyle[twoside,fleqn,espcrc2]{article} 
\newcommand{\AmS}{{\protect\the\textfont2
  A\kern-.1667em\lower.5ex\hbox{M}\kern-.125emS}} 
\newcommand{\beq}{\begin{equation}}
\newcommand{\eeq}{\end{equation}}
\newcommand{\bea}{\begin{eqnarray}}
\newcommand{\eea}{\end{eqnarray}}

\newcommand{\tr}{{\rm tr}}
\newcommand{\V}{{\cal V}}
\newcommand{\vev}[1]{\Big\langle #1 \Big\rangle}
\input epsf

\title{$SU(3)$ string tension and the presence of vortices}

\author{Tam\'as G. Kov\'acs\address{Department of Physics, 
        University of Colorado\\ Boulder, CO 80309-0390, USA}
        \thanks{Research partially supported by NSF PHY-9023257, 
         DE-FG02-92ER-40672} 
        and 
        E. T. Tomboulis\address{Department of Physics, 
        UCLA\\ 
        Los Angeles, CA 90095-1547, USA}
        \thanks{Research partially supported by NSF 
         PHY-95310223.}}
       
\begin{document}

\begin{abstract}
Lattice simulations are presented showing the expectation of the 
fluctuation of the Wilson loop solely by elements of the center to fully reproduce the SU(3) heavy quark potential. The results are stable under smoothing, and point to thick vortices as being responsible for the full 
SU(3) tension. An analytic result on the necessary presence of thick 
vortices for confinement at weak coupling is also presented.     
\end{abstract}   

\maketitle

Recently, substantial numerical evidence has been obtained 
for the vortex picture of confinement \cite{KT1} - \cite{S}. 
Here we present simulations extending our $SU(2)$ 
results in \cite{KT1} to the $SU(3)$ gauge group. 
We also present an analytical result \cite{KT2} on the necessity 
of the presence of thick vortices for maintaining confinement at 
weak coupling.

Recall that vortices are characterized 
by multivalued singular gauge transformations $V(x)\in SU(N)$  
with multivaluedness in the center $Z(N)$. Vortex 
configurations have a pure-gauge asymptotic tail 
given by $V(x)$ providing  
the topological characterization of the configurations 
irrespective of the detailed structure of the core. Such a 
$V(x)$, if extended throughout spacetime, becomes singular on  a 
closed surface $\V$ of codimension 2 forming the topological obstruction 
to a single-valued extension. 
On the lattice, the surface $\V$ of codimension 2 is regulated 
to a coclosed set $\V$ of 
plaquettes, i.e. a closed loop 
of dual bonds in $d=3$; a closed 
2-dimensional surface of dual plaquettes in $d=4$, 
and so on. This represents the core of a thin vortex, each 
plaquette in $\V$ carrying flux $z\in Z(N)$. 
Such a thin vortex is suppressed at large $\beta$ with a 
cost proportional to the size of $\V$. 

Thick vortex configurations can be constructed by
perturbing the bond variables $U_b$ in the boundary of each 
plaquette $p$ in $\V$ so as to cancel the flux $z$ on $p$, and 
distribute it over the neighboring plaquettes forming a thickened core. 
If $\V$ is extended enough, the core may be
made thick enough, so that each plaquette receives a 
correspondingly tiny portion of the original flux $z$ that 
used to be on each $p$ in $\V$. Long thick vortex configurations  may 
therefore be obtained having $\tr U_p\sim \tr1$ for all $p$ on 
$\Lambda$. Thus they may survive at weak coupling. 

Hybrid vortex configurations having a thick and a thin part 
are also possible. Such hybrid vortices 
formed by long thick vortices `punctured' by a short (e.g. 
one-plaquette-long) thin part may then also survive at weak coupling. 
The short thin part may serve as a `tagging' of a long thick vortex. 

Consider two gauge configurations  $\{U_b\}$ 
and $\{U_b^\prime\}$ that differ by such a singular gauge 
transformation $V(x)$, and let $U[C]$ and $U^\prime[C]$ 
denote the respective path ordered  products around a loop $C$.   
Then $\tr\,U^\prime[C]=z\,
\tr\,U[C]$, where $z\neq1$ is a nontrivial element of the center,
whenever $V$ has obstruction $\V$ linking with the loop 
$C$; otherwise, $z=1$. Conversely, changes in the value 
of $\tr U[C]$ by elements of the center can be undone by 
singular gauge transformations on the gauge field 
configuration linking with the loop $C$. This means that 
vortex configurations are topologically characterized by elements 
of $\pi_1(SU(N)/Z(N))=Z(N)$. Thus the fluctuation 
in the value of $\tr U[C]$ by elements of $Z(N)$ expresses 
the changes in the number (mod $N$) of vortices linked with the 
loop over the set of configurations for which it is evaluated.

Motivated by this picture, 
we separate out the $Z(N)$ part of the Wilson loop observable 
by writing $\arg (\tr U[C]) = \varphi[C]+ {2\pi \over N}n[C]$, 
where $-\pi/N < \varphi[C]\leq \pi/N$, and $n[C]=0,1,\ldots,N-1$. 
Thus, with $\eta[C]=\exp(i{2\pi \over N}n[C]) \in Z(N)$, 
\bea 
W[C]& = &\vev{\tr U[C]} 
           = \vev{|\tr U[C]|\,e^{i\varphi[C]}\,\eta[C]}\label{WL}\\
               & =& \vev{ |\tr U[C]|\,\cos(\varphi[C])\,
               \cos({2\pi \over N}n[C])} \label{WLdec}
\eea 
where the last equality follows from the fact that the 
expectation is real by reflection positivity, and that it is 
invariant under $n[C] \to (N-n[C])$.  We next define 
\beq 
W_{Z(N)}[C] = \vev{\cos({2\pi \over N}n[C])}\; \label{ZnWL}
\eeq 
for the expectation of the $Z(N)$ part, which, as noted above,  
gives the response to the fluctuation in the number (mod $N$) 
of vortices linking with the loop. In the following we 
compare the string tension extracted from the full Wilson loop 
$W[C]$, eq. (\ref{WL}), to the string tension extracted from 
$W_{Z(N)}[C]$, eq. (\ref{ZnWL}), for $N=3$. 
\begin{figure}[h]
\vspace{-1cm}
\begin{center}
\leavevmode
\epsfxsize=70mm
\epsfysize=70mm
\epsfbox{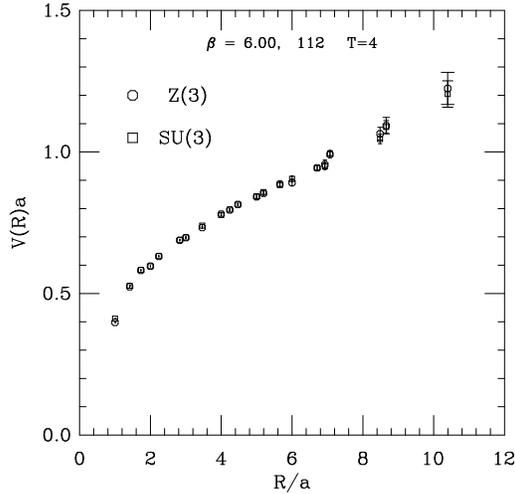}
\end{center}
\vspace{-1.2cm}
\caption{The heavy quark potential at $\beta=6.0$ on a set of
112 $12^3*16$ lattices.}
   \label{fig:potsu3_b6.0}
\vspace{-0.8cm}
\end{figure} 
\begin{figure}[htb]
\vspace{-1cm}
\begin{center}
\leavevmode
\epsfxsize=70mm
\epsfysize=70mm
\epsfbox{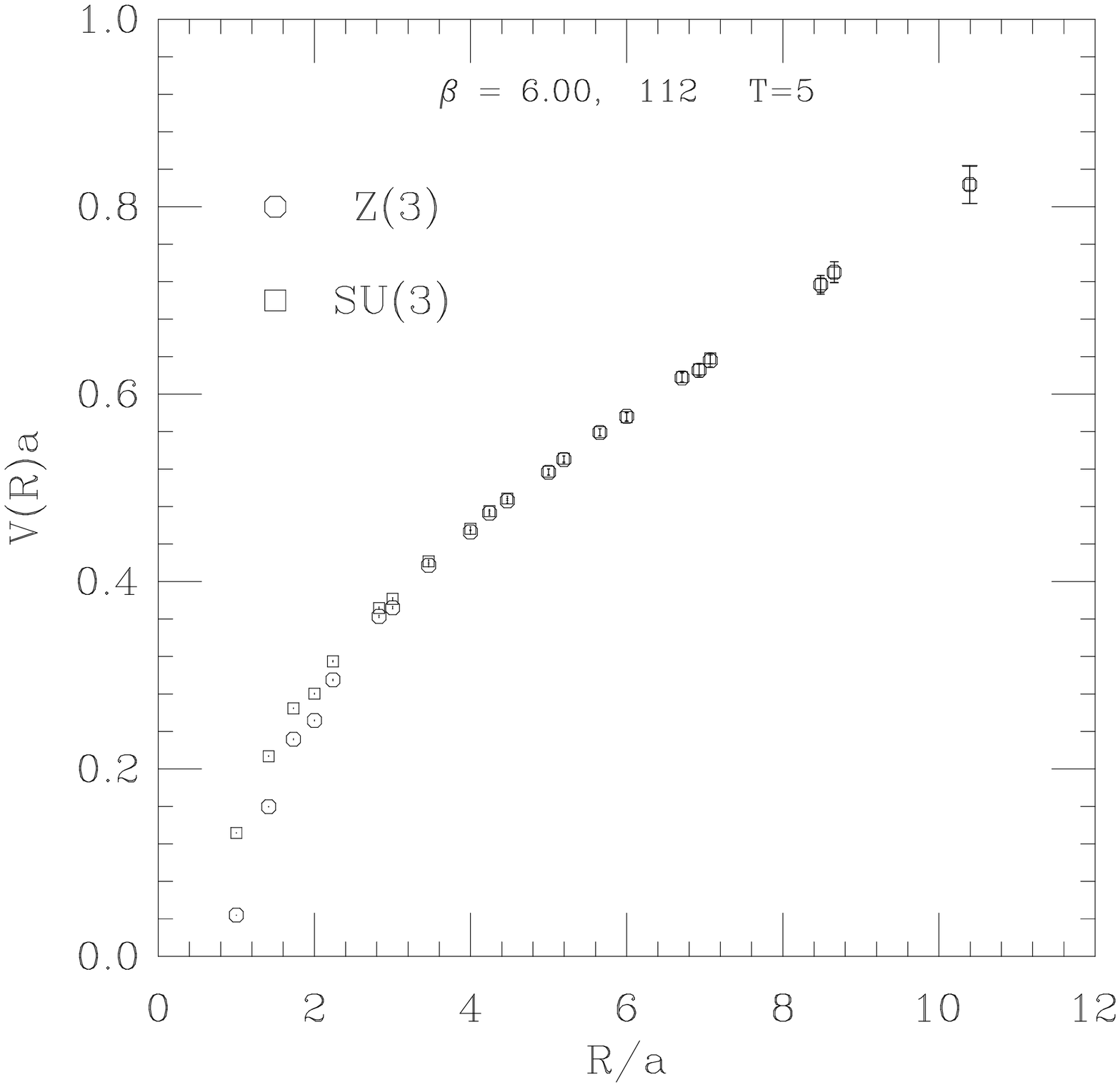}
\end{center}
\vspace{-1.2cm}
\caption{The heavy quark potential at $\beta=6.0$ on a set of
112 $12^3*16$ lattices from 2 times smoothed lattices.}
   \label{fig:potsu3_b6.0_b2}
\vspace{-0.8cm}
\end{figure}
\begin{figure}[htb]
\vspace{-1cm}
\begin{center}
\leavevmode
\epsfxsize=70mm
\epsfysize=70mm
\epsfbox{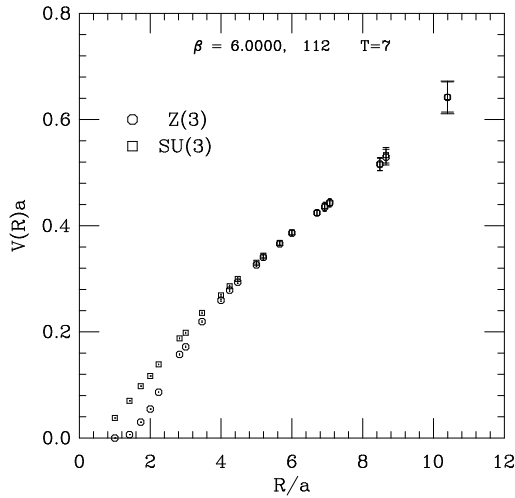}  
\end{center}
\vspace{-1.2cm}
\caption{The heavy quark potential at $\beta=6.0$ on a set of
112 $12^3*16$ lattices from 6 times smoothed lattices.}
   \label{fig:potsu3_b6.0_b6_t4}
\vspace{-0.8cm}
\end{figure}
\ \\

We worked with the Wilson action at lattice spacings 
$a=0.15$ fm and $a=0.10$ fm for $\beta=5.8$ and $\beta=6.0$, 
respectively. This is computed from the string tension
assuming that its physical value is 440MeV.

Results are presented in Fig.  
\ref{fig:potsu3_b6.0}. The agreement between the potential   
extracted from the full Wilson loop and that from the $Z(3)$ 
fluctuation expectation (\ref{ZnWL}) is striking. Note that it 
includes also the short-distance regime. This is because  
(\ref{ZnWL}) counts {\it both} thick and thin 
vortices, and the thin ones are clearly important at short distances 
(narrow loops). At longer distances, 
however, only sufficiently thick vortices can be expected to 
contribute to the string tension.

To explore this we performed local smoothing on our 
configurations which removes short distance fluctuations but 
preserves the long distance physical features. 
It should be noted that this is in fact essential for checking that 
topological structure on the lattice is actually well represented.  
According to well-known results, unless the variation of a lattice 
configuration over short distance is restricted enough, there is no 
unambiguous connection to a continuum extrapolation and hence assignment 
of topology.  
If the string tension is then really fully reproduced by the vortex 
fluctuations, the agreement seen in Fig. 
\ref{fig:potsu3_b6.0} should persist 
at long distances when the potential is measured  
on the smoothed configurations. 
This is indeed a very stringent test.  
We used the smoothing procedure of Ref. \cite{DeGet} applied 
here to $SU(3)$. Results for the potentials on 
twice smoothed configurations are given in 
Fig. \ref{fig:potsu3_b6.0_b2}.     

We see that the potentials extracted from $W[C]$ and 
$W_{Z(3)}[C]$ now disagree over short distances, but then again 
merge together with no discernible difference at distances 
$R/a > 3$. This is as expected: smoothing destroys thin 
vortices but leaves vortices thicker than the smoothing 
scale unaffected.

Performing successive smoothing steps extends the 
distance scale over which fluctuations are smoothed; but the 
asymptotic string tension should not be affected, since, for 
sufficiently large loops, there is a scale beyond which linked 
thick vortices are not affected. This is clearly illustrated by 
comparing Fig. \ref{fig:potsu3_b6.0_b2} to 
Fig. \ref{fig:potsu3_b6.0_b6_t4} which displays the potentials 
resulting on six times smoothed configurations. 

What if, on the other hand, we eliminate all thick 
vortices linked with the loop, 
but leave {\it thin} vortices intact?  A very convenient formalism, which  
explicitly separates thin and thick vortices, recasts the $SU(N)$ 
theory in a $Z(N)\times SU(N)/Z(N)$ form \cite{KT1}. Thin vortices are 
then described purely in terms of $Z(N)$ variables. 
In this language it is straightforward to modify the Wilson loop operator 
by a constraint that forbids its fluctuation by center elements due 
to linkage with thick vortices. We should then expect the  
string tension to actually vanish. In fact, we have obtained an 
analytic  proof of this fact. For $SU(2)$ we show \cite{KT2} that:\\
For sufficiently large $\beta$, and dimension 
$d\geq 3$ the so 
constrained Wilson loop expectation 
exhibits perimeter law, i.e. 
there exist constants $\alpha,\ \alpha_1(d),\ \alpha_2(d)$ 
such that  
\beq 
W[C] \,\geq \, \alpha \exp\left(\, - \alpha_2\,e^{-
\alpha_1\beta}\,|C|\,\right) \quad.\label{len}
\eeq 
Here $|C|$ denotes the perimeter length of the loop $C$. 
In other words, the potential indeed becomes nonconfining at weak 
coupling.  

In conclusion, all our results are   
consistent with a physical picture of locally smooth extended 
thick vortices occurring over all large scales, and giving rise 
to the full asymptotic string tension. As mentioned above, individual 
thick vortices may be `tagged' by the insertion of a short thin segment. 
We have proposed in the past that such tagging may be used to  
estimate the contribution of vortices more directly. Ongoing work in this 
direction will be reported elsewhere.

\end{document}